\documentclass[twocolumn,superscriptaddress,pra,amsmath,amssymb]{revtex4}
\usepackage{graphicx}

\newcommand{\la}{\langle}
\newcommand{\ra}{\rangle}
\newcommand{\diff}{\mathrm{d}}
\newcommand{\sinc}{\text{\rm sinc}}

\newcommand{\V}{ {\cal V} }

\begin{document}

\title{Absolute absorption spectroscopy based on molecule interferometry}
\author{Stefan Nimmrichter}%
\affiliation{Faculty of Physics, University of Vienna, Boltzmanngasse 5, 1090 Vienna, Austria}

\author{Klaus Hornberger}%
\affiliation{Arnold Sommerfeld Center for Theoretical Physics,
Ludwig-Maximilians-Universit\"{a}t M\"{u}nchen,\\
Theresienstra{\ss}e 37, 80333 Munich, Germany }

\author{Hendrik Ulbricht}%
\affiliation{Faculty of Physics, University of Vienna, Boltzmanngasse 5, 1090 Vienna, Austria}

\author{Markus Arndt}%
\affiliation{Faculty of Physics, University of Vienna,
Boltzmanngasse 5, 1090 Vienna, Austria}

\date{\today}

\begin{abstract}
We propose a new method to measure the absolute photon absorption
cross section of neutral molecules in a molecular beam. It is
independent of our knowledge of the particle beam density, nor does
it rely on photo-induced fragmentation or ionization.  The method is
based on resolving the recoil resulting from photon absorption by
means of near-field matter-wave interference, and it thus applies
even to very dilute beams with low optical densities. Our discussion
includes the possibility of internal state conversion as well as
fluorescence. We assess the influence of various experimental
uncertainties and show that the measurement of absolute absorption
cross sections is conceivable with high precision and using existing
technologies.
\end{abstract}

\maketitle

\section{Introduction}
Optical spectroscopy is a key technology for unraveling electronic,
vibrational, rotational and structural properties of atoms,
molecules and clusters with applications in physics, astronomy,
chemistry, and biology. Specifically, their absorption cross
sections are nowadays routinely measured with commercial
spectrometers, using a large number of precise and sensitive methods
covered in modern
textbooks~\cite{Scoles1988a,Kreibig1995a,Berkowitz2002a}. A general
interest in these applications is to improve the spectral
resolution, the sensitivity to small sample quantities, and the
precision with regard to the absolute magnitude of the absorption
cross section.

In the present article, we propose  a new method for determining the
{\em absolute } value of the absorption cross section of neutral
nanoparticles at a given optical frequency. In molecular or cluster
beam physics this is usually a non-trivial task: Although absolute
absorption {\em coefficients} can be measured rather precisely even
at low densities,  the extraction of absorption {\em cross sections}
is often limited to volatile substances, whose pressure and number
density can be determined with sufficient
accuracy~\cite{Sakurai1975a,Chen1999a}. For complex nanoparticles,
information about the particle number is often only available with
low precision. Instead, some experiments occasionally use
information about the absorption cross section to characterize the
supersonic beam composition~\cite{Boogaarts1995a}. Even for effusive
beams, which are better controlled and understood, inaccuracies in
measuring the source pressure often lead to a large error in the
derived optical cross section~\cite{Coheur1996a,Yasumatsu1996a}.

In all these cases, optical spectra may still be recorded with very
high spectral resolution using established methods, yet to properly
normalize the entire spectrum it remains important to determine the
absolute absorption cross section at least at one anchor point,
i.e., at a single wavelength. Such a measurement of an absolute
cross section can be achieved by referencing the absorption to a
different signal that is derived from the same molecular beam. In
the past, various such in-situ normalization methods have been
successfully developed. For instance, absolute spectra of metal
clusters were determined by the method of photo-depletion, where a
photon absorption suffices to remove the cluster from the observed
spectrum~\cite{Wang1990a,Knickelbein1992a}. Similarly, the rate of
photo-fragmentation of poly-aromatic hydrocarbons~\cite{Pino1999a}
or the survival time of fullerene dianions in a penning
trap~\cite{Concina2008a} was successfully used for determining
absolute photo absorption cross sections.

However, the mentioned techniques share one potential drawback:
they rely on photo-induced changes of the particle composition.
This may be undesirable in some situations, and it can involve ambiguities. In
particular, for mass mixtures of clusters or large polymers it may occur that the
mass spectrum is photo-depleted towards smaller masses but
simultaneously refilled from high mass
compounds~\cite{Brockhaus1999a}. Although recently
established transition matrix methods allow to handle rather
complex decay chains~\cite{VanOanh2006a}, it still appears
desirable to develop a method that permits to determine absolute
absorption cross sections without modifying the molecule and
without knowing the particle density.

The measurement scheme proposed below is based on the insight that
the least invasive spectroscopic method consists in determining the
momentum recoil imparted by a single photon on the flying molecule.
While single-photon recoil effects have already been proposed for
non-linear spectroscopy of atoms~\cite{Guo1992a}, the photon
momentum is usually significantly smaller than the molecular
momentum spread in those classical beam experiments, so that the
resulting beam broadening is hard to resolve. Modern quantum
interferometers show a way around this limitation, and, indeed,
atom-interferometric recoil measurements have already led to
precision experiments. For instance, the photon-induced recoil of
neutral alkali atoms allows one
 to determine the value of $h/m_{\rm atom}$, which enters in an
improved measurement of the fine structure constant
$\alpha$~\cite{Weiss1993a,Gupta2002a,Clade2006a}. Optical momentum
transfer was also central in investigations of photon-induced
decoherence processes in matter wave
interferometry~\cite{Pfau1994a,Chapman1995a}, and the single-photon
recoil was an essential part in Ramsey-Bord\'{e} interferometry with
I$_2$~\cite{Borde1994a}.

In the present contribution, we aim at an interferometric
determination of the absolute photon absorption probability through
a comparison of undeflected and  deflected molecular interference
patterns. We will see that near-field matter wave interferometry is
particularly well suited for this purpose, since it imposes a
spatial nanostructure on the traversing molecular
beam~\cite{Berninger2007a,Hackermuller2007a} and since the
interferometer fringe visibility allows us to precisely determine
any changes caused by single photon recoil. After outlining the
general idea, we will present a quantum theoretical description of
the effect. It is sufficiently detailed to permit assessing the
expected measurement accuracy in case of the photon absorption
alone. In a second step, we will also account for the possible
effect of fluorescence, resulting in a precise measurement scheme
for both the absolute absorption cross section and for the quantum
yield of fluorescence of neutral complex molecules and
nanoparticles.

\section{Basic setup}
\begin{figure}
\includegraphics[width=\columnwidth]{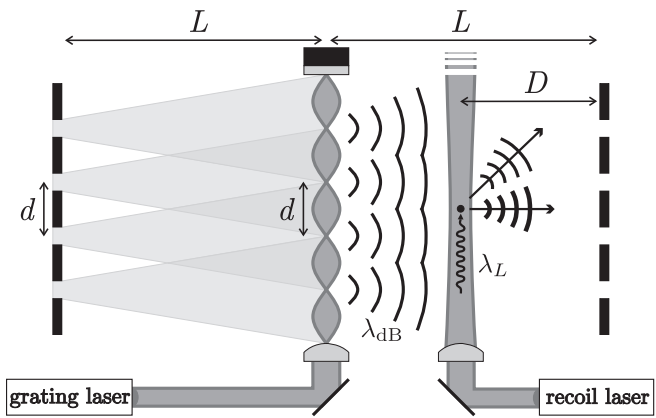}
\caption{\label{fig:setup} Setup of the proposed measurement scheme.
The molecular beam enters from the left through a material grating,
and the transmitted molecules are counted after passing the grating
mask on the right hand side. In  the center, a standing light field
created by the grating laser serves as a diffraction element for
near-field Talbot-Lau interference of the molecular matter waves.
When the molecules pass the recoil laser beam they may absorb an
integer number of photons, thus effecting a probabilistic
modification of the interference pattern. For an appropriate choice
of the distance $D$, the nanostructured, periodic form of the
interference pattern thus enables a robust and precise extraction of
the molecular absorption cross section by observing how of the
quantum interference contrast reduces as a function of the recoil
laser power.}
\end{figure}

The measurement scheme is based on the Talbot Lau interferometer
(TLI), which consists of three gratings, as sketched in
Fig.~\ref{fig:setup}. The first and the third one are material
masks, while the second diffraction element can be either an
absorptive material \cite{Brezger2002a} or a pure phase grating
created by a standing laser wave. The latter variant, designated as
the Kapitza-Dirac-Talbot-Lau configuration (KDTLI)
\cite{Gerlich2007a}, has the advantage that the power of the optical
diffraction grating can be easily tuned, adding an additional handle
for the experiments~\cite{Gerlich2008a}.

In the basic setup the gratings have identical grating periods $d$
and are placed at equal distances $L$. A spatially incoherent, but
velocity-selected molecular beam with de Broglie wave length
$\lambda_{\rm dB}$ illuminates the first grating, creating the
spatial coherence for diffraction at the second grating. A near
field interference effect, which depends on the ratio of $L$ to the
Talbot length $L_{\rm T}=d^2/\lambda_{\rm dB}$ \cite{Talbot1836a},
yields a periodic density pattern further downstream. The
interference contrast depends on the  wave length of the molecules,
determined by their longitudinal velocity $v_z$, and also on the
specific interactions with  the diffraction structure.

The third grating serves as a transversally movable mask for a
detector that measures the total molecule flux behind the gratings.
The resulting fringe pattern as a function of the third grating
displacement characterizes the interference pattern. Its functional
form is well approximated, in both interferometer designs and under
many experimental circumstances, by a simple sine curve with an
offset.

We here propose to extend this arrangement by an additional recoil
laser beam, running parallel to the  gratings and crossing the
molecular beam perpendicularly at a variable position $D$ between
the second and the third grating, as sketched in
Fig.~\ref{fig:setup}. Depending on the recoil laser power $P_L$, its
wavelength $\lambda_L$, and the molecular absorption cross section
$\sigma_\text{abs}$, each molecule may absorb one or more photons or
may remain unaffected. The resulting recoils  shift the associated
interference patterns by an amount which depends on the distance $D$
and on the molecular velocity $v_z$. The observed fringe pattern is
an average of all individual molecular interference patterns, whose
weights are determined by the absorption cross section. We will show
in the following that the observed fringe visibility, and in
particular its variation with the spectroscopy laser power, then
allow one to derive a precise value for the absolute optical
absorption cross section.

\section{Theoretical analysis}

The motional state of the molecules is most conveniently
represented in phase space, where it is characterized by the Wigner
function $w(x,p)$ of the transverse position $x$ and momentum $p$
with respect to a longitudinally co-moving frame at fixed beam
velocity $v_z$ \cite{Hornberger2004a,Nimmrichter2008a}. In this
framework, the absorption of each individual laser photon leads to a
momentum kick $\Delta p = h/\lambda_L$. The resulting fringe
pattern at the third grating is then shifted by
\begin{equation}
s = \frac{\Delta p}{p_z} D =  \frac{\lambda_\text{dB}}{\lambda_L} D,
\label{eqn:s}
\end{equation}
where $\lambda_\text{dB} = h / mv_z$ is the de Broglie wavelength of
the molecules. We require that the shift $s$ be less than one
grating period $d$ in order to maintain an unambiguous assignment
between the fringe shift and the absorption cross section. This
limits the range of possible spectroscopy wavelengths $\lambda_L$ to
$\lambda_L \leq \lambda_\text{dB} L / d$, since $D<L$ (which is a
minor constraint since a single wavelength is often sufficient for
calibrating an entire spectrum).

The spatial density distribution at the third grating is given by
an incoherent sum of the shifted distributions,
\begin{equation}
    w_3(x;n_0,s) = \sum_{n=0}^{\infty} P_n (n_0) w_3^{(0)} \left(x + n s \right) \label{eqn:w3shift},
\end{equation}
where $w_3^{(0)}(x)$ denotes the periodic density pattern without the recoil laser.
The probability $P_n (n_0)$ that a molecule absorbs $n$ photons
while crossing the laser can be modeled by a Poissonian
distribution, $P_n (n_0) = {n_0^n} \exp({-n_0})/{n!}$. The average
number of absorbed photons $n_0$ is obtained by integrating the
recoil laser intensity  over the longitudinal Gaussian laser profile,
\begin{equation}
    n_0 = \sqrt{\frac{2}{\pi}}\frac{\sigma_\text{abs} P_L \lambda_L}{h c w_y v_z}, \label{eqn:n0}
\end{equation}
where $w_y$ is the laser waist perpendicular to the molecular
beam and $\sigma_\text{abs}$ the absorption cross section of the
molecules. Here the transverse motion of the molecules through the beam can
be safely neglected since the molecular beam is typically collimated to
about 1\,mrad~\cite{Hornberger2008b}.

The assumption of a Poissonian distribution $P_n (n_0)$ is valid if
consecutive absorption processes are independent, i.e. if the cross
section $\sigma_\text{abs}$ does not change after the first
excitation. In general, this cannot be taken for granted. In fact,
various cases of optical limiting are known, with fullerenes as a
prominent example~\cite{Wray1994a}. In practice, the spectroscopy
laser power shall therefore always be limited to $n_0 < 1$, leading
to relaxed power requirements in the experiment.

The modified molecular density distribution (\ref{eqn:w3shift})
 is determined in Fourier space  by the coefficients
\begin{equation}
    w_\ell (n_0,s) = w_\ell^{(0)} \exp \left( n_0 \exp \left( \frac{2\pi i \ell s}{d} \right) - n_0 \right)\,,
 \label{eqn:wlshift}
\end{equation}
with $w_\ell^{(0)}$ the Fourier series coefficients  of the
unperturbed interference pattern $w_3^{(0)}(x) = \sum_\ell
w_\ell^{(0)} \exp (2\pi i \ell x / d)$. Since the real part of the
exponent in (\ref{eqn:wlshift}) is negative for $\ell \neq 0$ the
contrast of the density pattern decreases if a recoil laser is
applied, i.e. the Talbot-Lau interference blurs. This effect gets
less important when the recoil laser approaches the third grating.
Note that the unperturbed coefficients $w_\ell^{(0)}$, the recoil
shift $s$, and the mean absorption $n_0$ depend on the molecular
velocity, taken to be at a fixed value $v_z$ for the time being. A
finite velocity spread in the beam can be accounted for in the end
by taking the average over the molecular velocity distribution.

The detected signal $S(x)$ as a function of the third grating
displacement $x$ is subject to the same transformation
(\ref{eqn:wlshift}), since its Fourier coefficients $S_\ell$ differ
from the $w_\ell$ by a multiplication with the coefficients of the
transmission function of the third grating
\cite{Hornberger2004a,Nimmrichter2008a}. Our treatment assumes that
the detection process is independent of the internal molecular
state, which is well justified since the detection process behind
the interferometer typically occurs several hundred microseconds
after the excitation inside the interferometer. The incident photon
energy will have been either re-radiated in a fluorescence process,
or will have been dissipated into many vibrational modes to which
most detection mechanisms are insensitive.

It is interesting to note that one could even measure  separately a
possible absorption-induced change in the detection efficiency, by
choosing the distance $D$ such as to set the photon-induced shift to
equal a full fringe period $s = k d$. Since this alone does not
change the experimental signal, additional effects related either to
fluorescence effects or to photon-induced modifications of the
detection efficiency thus get experimentally accessible.

The effect of the laser recoil is maximized, on the other hand,
when the density pattern is shifted by half its period $d$ per
photon, i.e. by $s= (2k+1)d/2$ with integer $k$ for a fixed
molecular velocity $v_z$. Equation (\ref{eqn:w3shift}) then simplifies
to
\begin{eqnarray}
    w_3 \left(x;n_0, \frac{d}{2} \right) &=& e^{-n_0} \cosh (n_0) \Big[ w_3^{(0)}\left(x\right)
\nonumber \\
&& +\tanh (n_0) w_3^{(0)}\Big( x  + \frac{d}{2} \Big) \Big]
\label{eqn:halfperiod},
\end{eqnarray}
where the ratio of the shifted and the unshifted pattern, given by $\tanh
(n_0)$, rapidly approaches unity for $n_0 > 1$.
\begin{figure}
\includegraphics[width=8cm]{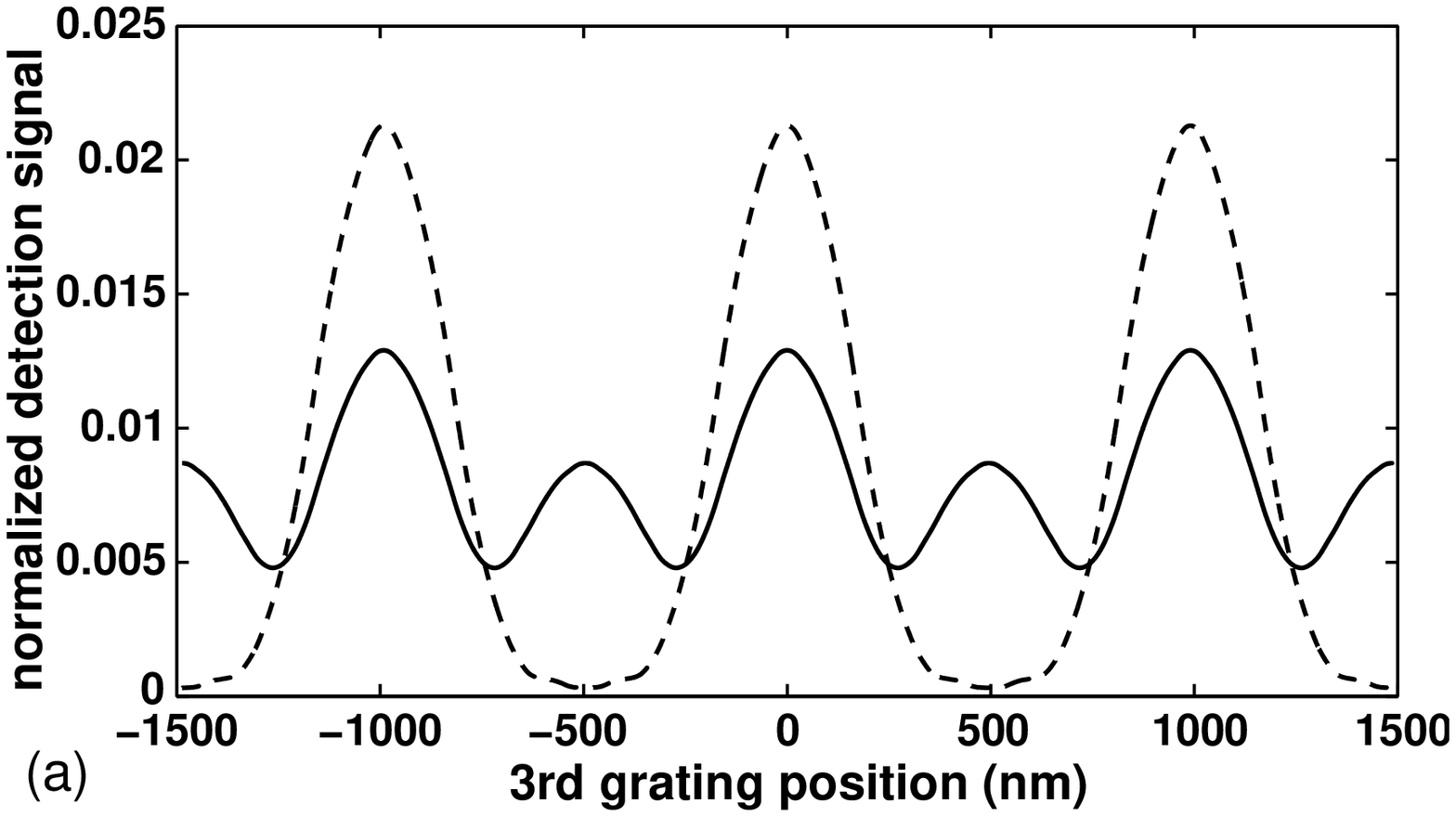}\\
\vspace{3mm}\includegraphics[width=8cm]{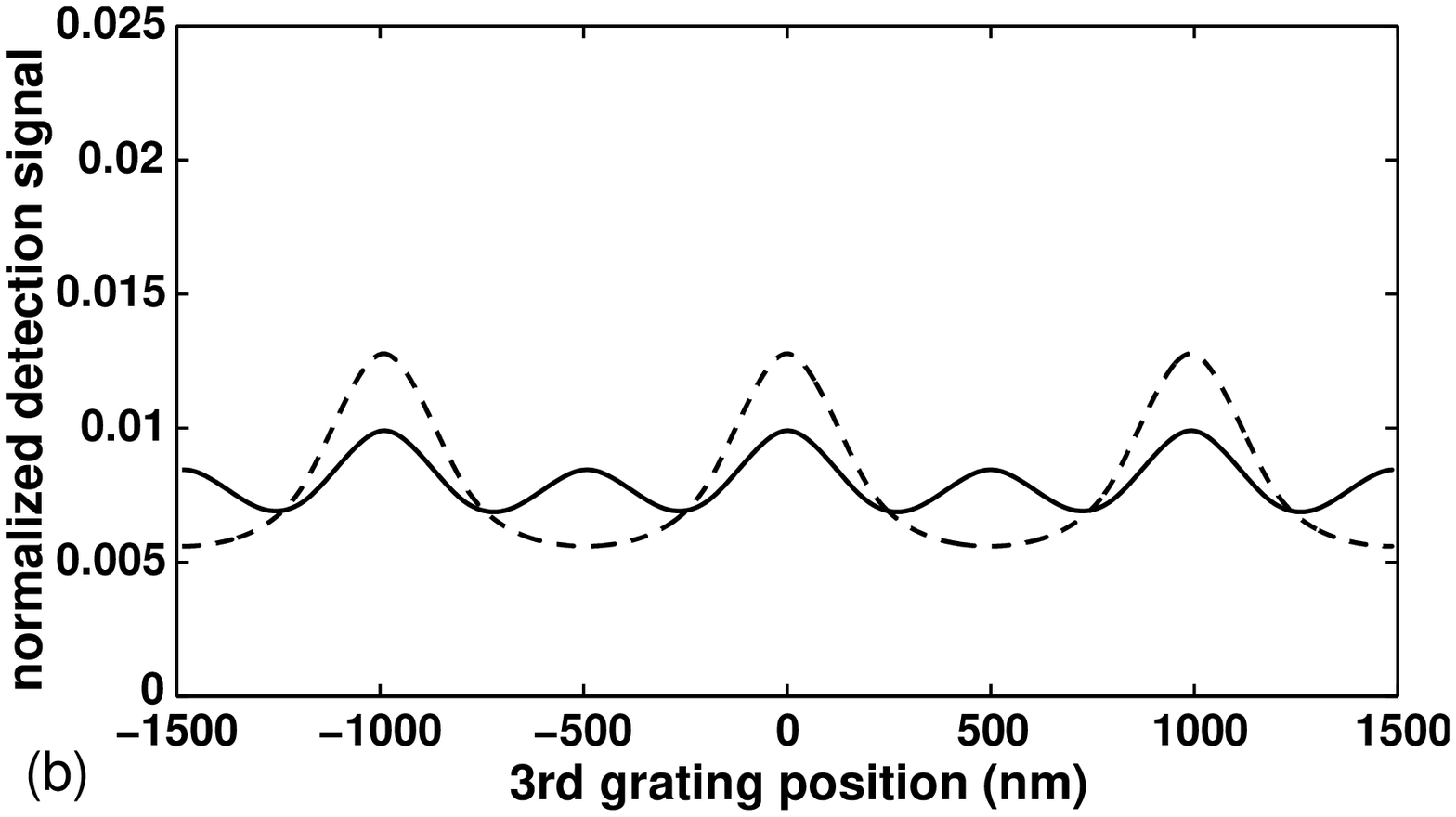}
\caption{\label{fig:doppelpeak} Effect of a half-period recoil
shift on the interference signal of a TLI setup composed of three
metallic gratings with a period of $991\,$nm and an open fraction
of f=1/5, implying a slit width of $a=198.2\,$nm. The plots show the
expected signal as a function of the transverse position of the
third grating, normalized to the flux of the molecular beam
entering the interferometer. The dashed line in panel (a) shows the peaked
fringe pattern for tetraphenylporphyrin (H$_2$TPP) molecules
passing the interferometer at fixed velocity of $175\,$m/s. It is
changed to the solid line by a $420\,$nm recoil laser of $0.5\,$W
power that shifts the pattern by half a period per absorbed
photon. Panel (b) shows an analogous situation, where the molecular
beam displays a Gaussian velocity distribution centered at
$\overline{v}_z=175\,$m/s with a relative width of $\Delta v/\overline{v}_z=5\,\%$.}
\end{figure}

This special case is illustrated in Fig.~\ref{fig:doppelpeak}, for a
standard TLI setup with three equidistant material gratings of
period $d=991\,$nm and an open slit fraction $f=a/d=0.2$. Here and
in the following examples we use the biodye tetraphenylporphyrin
H$_2$TPP at an average velocity $\overline{v}_z=175\,$m/s. It has a
mass of $m=614.74\,$amu, a static polarizability of
$\alpha=105\,$\AA$^3$ \cite{Deachapunya2007b}, and it exhibits a
large optical absorption cross section
$\sigma_\text{abs}=15\,$\AA$^2$ at the resonance frequency
$\lambda_L=420\,$nm \cite{Du1998a}. This corresponds on average to
$n_0=1.6$ absorbed photons per Watt power of the spectroscopy laser
at that wavelength, when it is focused on a waist of
$w_y=900\,\mu$m, which is typical for current interferometry
experiments~\cite{Gerlich2007a}.

Figure~\ref{fig:doppelpeak} shows the expected signal
as a function of the third grating
position when the grating distance matches exactly the Talbot
length $L_\text{T}=m v_z d^2 /h=26.48\,$cm. Due
to the small grating slits and the fixed velocity $v_z=175\,$m/s
in Fig.~\ref{fig:doppelpeak}a, one observes a fringe pattern with
distinct narrow peaks (dashed line). Photo absorption from a
laser of $P=0.5\,$W positioned at $D=5.61\,$cm, generates growing
side peaks shifted by half a fringe distance and reduces the
magnitude of the original peak accordingly (solid line), as described by
Eq.~(\ref{eqn:halfperiod}).

The appearance of a distinct side-peak can be used as a clear
indication of the recoil effect. Absolute cross sections can then be
determined by comparing the side peak height to that of the original
peak using Eq.~(\ref{eqn:halfperiod}). This concept should work very
well for molecular beam sources that can be made sufficiently
intense to permit a velocity selection well below one percent. This
holds in particular for experiments that exploit laser desorption
into a freely expanding seed gas jet~\cite{Marksteiner2007a}.

A precise quantitative evaluation may, however, often be compromised
in practice by a finite velocity spread $\Delta v_z/\overline{v}_z$
of the molecule beam, as demonstrated in Fig.~\ref{fig:doppelpeak}b.
It assumes a Gaussian velocity distribution  centered at
$\overline{v}_z=175\,$m/s with a relative width of $5\,\%$. The
interference pattern gets blurred  because of the strong velocity
dependence of the interference contrast of a TLI with small-slit
material gratings \cite{Gerlich2007a,Nimmrichter2008a}. The signal
peaks (dashed line) therefore overlap and render the extraction of
the absorption cross section inaccurate. A growing molecular mass
and polarizability, as well as the possibility of fluorescence, will
worsen the situation. Therefore, when precision and scalability to
larger molecules matter, a tunable KDTLI laser grating setup is to
be preferred to the TLI,  even though the qualitative distinction of
two separate sets of interference peaks gets lost.

Specifically, in case of the near-sinusoidal fringe patterns expected in a
KDTLI setup \cite{Gerlich2007a} the recoil laser simply reduces
the detected sinusoidal fringe visibility $\V=2|S_1/S_0|$ by the
factor
\begin{equation}
    R(n_0,s) = \exp \left( - n_0 \left[ 1 - \cos \left( \frac{2\pi s}{d}\right) \right] \right), \label{eqn:R1}
\end{equation}
according to Eq.~(\ref{eqn:wlshift}). The reduction is mainly
determined by the mean absorbed photon number $n_0$ and it can be
at most $ R(n_0,(2k+1)d/2) = \exp \left(-2n_0 \right)$ in case of a
half-period shift. It follows with Eq.~(\ref{eqn:n0}) that a linear relation
holds between the logarithm of $R$ and the cross section
$\sigma_\text{abs}$,%
\begin{equation}
     \ln R^{-1} = 2 \sqrt{\frac{2}{\pi}}\frac{\lambda_L}{h c w_y v_z} \sigma_\text{abs} P_L
\, . \label{eqn:lnR1}
\end{equation}
One therefore easily extracts $\sigma_\text{abs}$
from a linear regression to the measured contrast reduction for
different recoil laser powers $P_L$. This procedure is rather
robust against dephasing and contrast-reducing effects inside the
interferometer, such as grating vibrations, gravitational and
Coriolis forces~\cite{Stibor2005a}, collisions with residual gas
molecules~\cite{Hornberger2003a} or the absorption of photons
at the second laser grating
\cite{Gerlich2007a,Hornberger2008b}, since the only relevant effect
is the reduction of the fringe contrast \textit{relative} to the
value in absence of the recoil laser.

However, this method still displays some sensitivity to the velocity
spread of a realistic molecule beam. Not only the interference
pattern itself  depends on the longitudinal velocity $v_z$, but also
the mean absorbed photon number (\ref{eqn:n0}), as well as the
fringe shift (\ref{eqn:s}) imparted by the recoil laser. Each
Fourier component (\ref{eqn:wlshift}) of the interference pattern
must therefore be averaged over a realistic velocity distribution
$\mu (v_z)$. This  affects the terms $w_\ell^{(0)}$, $n_0$ and $s$.
The reduction factor (\ref{eqn:R1}) is then replaced by
\begin{equation}
    \la R\ra_{v_z} = \left| \frac{\la w_1^{(0)} \exp \left( - n_0 \left[ 1 - \exp \left( \frac{2\pi i s}{d}\right) \right] \right) \ra_{v_z}}{\la w_1^{(0)} \ra_{v_z}} \right|. \label{eqn:R2}
\end{equation}
If this was used as a fitting expression to extract the cross
section from the contrast reduction it would again require knowledge
of the exact velocity distribution. However, this complication can
be avoided even for moderate velocity spreads, as we will discuss in
the following.

\section{Expected accuracy}

In our setup the molecular velocity spread is a main cause for
uncertainties when extracting the absorption cross section from
Eq.~(\ref{eqn:lnR1}). In order to estimate the influence of the
velocity spread we assume a symmetric velocity distribution $\mu
(v_z)$ peaked at the value $\overline{v}_z$ with a characteristic
width of $ \Delta v_z$. The recoil laser is set such that the mean
recoil shift amounts to one half of the grating period at
$\overline{v}_z$, $\overline{s}=d/2$. It is not advisable to use
larger shifts $\overline{s}=nd+d/2$, with integer $n$, because of
their correspondingly larger variation due to the velocity spread.

For a sufficiently small relative width $\Delta v_z / \overline{v}_z$ one may
expand the full expression (\ref{eqn:R2}) for the contrast
reduction up to quadratic order in $\Delta v_z / \overline{v}_z$
and carry out the integrations to obtain
\begin{equation}
 \frac{\Delta R}{\la R \ra_{v_z}} \approx 2\overline{n}_0 \left(\frac{\Delta v_z}{\overline{v}_z}\right)^2 \left( \overline{n}_0 - 1 + \frac{\pi^2}{2} + \frac{\overline{v}_z}{\overline{\V}_0} \frac{\partial \overline{\V}_0}{\partial \overline{v}_z} \right)\label{eqn:R3},
\end{equation}
with $\overline{\V}_0$ the visibility in absence of the recoil
laser and $\overline{n}_0$ the mean absorbed photon number at
$\overline{v}_z$. The influence of the velocity dependence  of the
interference contrast $\overline{\V}_0$ is naturally suppressed by
choosing  setup parameters that maximize its value
with $\partial \overline{\V}_0 / \partial \overline{v}_z = 0$.
This only holds as long as the contrast varies slowly
within $\overline{v}_z \pm \Delta v_z$, which should be
experimentally achievable for relative velocity spreads below
$10\,\%$ ~\cite{Gerlich2007a,Nimmrichter2008a}. The remaining
uncertainty in (\ref{eqn:R3}) results from the velocity dependence
of the photon number $n_0$ and the recoil shift $s$. Since $n_0
\lesssim 1$ the relative error related to
the velocity distribution is no more than $\Delta R / \la R
\ra_{v_z} = \pi^2 \overline{n}_0 \Delta v_z^2 /
\overline{v}_z^2$, which directly contributes to the systematic
error $\Delta \sigma_\text{abs} / \sigma_\text{abs}$ of the derived absorption cross
section, according to Eq.~(\ref{eqn:lnR1}).

With a fair velocity selection of about $1\%$, which should be
feasible in future experiments ~\cite{Marksteiner2007a}, the error
can thus be reduced to $0.1\%$. Other experimental uncertainties,
such as the  power $P_L$ and waist of the recoil laser or its
distance $D$ to the third grating, can also be reduced to the
percent level.

\begin{figure}
\includegraphics[width=8cm]{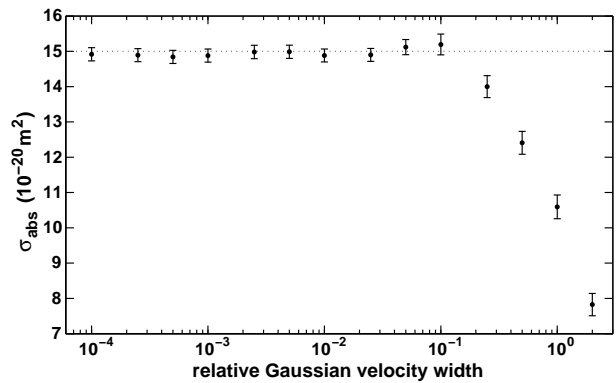}
\caption{\label{fig:sigmafit} Monte Carlo simulation of the
proposed measurement of the absorption cross section of H$_2$TPP
at $420\,$nm (see text). The results (including error bars at a
$95\,\%$ confidence level) are plotted versus the relative width
$\Delta v_z / \overline{v}_z$ of a Gaussian velocity distribution
of the molecules underlying the simulation. The dotted line marks
the true value of the absorption cross section
$\sigma_\text{abs}=15\,$\AA$^2$ \cite{Du1998a}. It is well reproduced by the simulated
measurement for velocity spreads below $10\,\%$.}
\end{figure}

The precision of our recoil spectroscopy method therefore hinges
on a good velocity selection. This is demonstrated in
in Fig.~\ref{fig:sigmafit} where we show the results of a
Monte Carlo simulation of a KDTLI experiment for H$_2$TPP
molecules. We use the setup parameters
from an existing experiment~\cite{Gerlich2007a} ($L=10.5\,$cm,
$d=266\,$nm) with a grating laser power of $8\,$W. In this situation the
velocity dependence of the interference contrast, i.e. $\partial
\overline{\V}_0 /
\partial \overline{v}_z$, can indeed be
neglected in Eq.~(\ref{eqn:R3}). The absorption of Laser grating
photons is included in the simulation, with a realistic cross
section of $0.17\,$\AA$^2$ \cite{Du1998a}. The $420\,$nm recoil
laser is placed at $D=1.51\,$cm, which conforms to a recoil shift
$s=d/2$ for molecules with $\overline{v}_z=175\,$m/s. Taking the
velocity distribution of the molecules to be a Gaussian centered at
$\overline{v}_z$, the simulation of the experiment is now done for
different relative widths of the distribution in
Fig.~\ref{fig:sigmafit}. Each data point represents the result of
fitting Eq.~(\ref{eqn:R1}) to a set of simulated visibilities, in
which the recoil laser power is ramped up to $1\,$W in $20$ steps.
In each step the raw statistical data is given by the number of
detected molecules as a function of the third grating position in a
total of $100$ samples over $5$ grating periods, where each sample
consists of $10^4$ molecules entering the KDTLI, a value in
accordance with the count rates obtained in recent experiments
\cite{Gerlich2007a}. A sine function is then fitted to the raw data for a
given recoil laser power to obtain the visibility
including a statistical error estimate.

The simulation corroborates the analytical estimate that the
absorption cross section (dashed line) could still be determined 
with a precision of about $1\,\%$ provided that the molecular
velocity spread can be reduced to below $10\,\%$. Above this value
the velocity dependence of the signal contrast gets significant and
the applied fit function (\ref{eqn:R1}) fails to reproduce the
correct value of the cross section. The error bars in
Fig.~\ref{fig:sigmafit} indicate the statistical uncertainty of the
fringe contrast, which is mainly due to the detection shot noise and
can be reduced by increasing the brilliance of the molecular beam.

The predicted accuracy of 1\% is valid if other experimental
parameters can be determined with that precision. This includes in
particular the waist, the power and the position of the spectroscopy laser beam. Using commercially available meters it is realistic to assume these parameters to be known to better than 5\%. On the one hand this accuracy would improve direct extinction measurements, where the restricted knowledge of the vapor pressure may yield uncertainties up to a factor of two \cite{Coheur1996a,Yasumatsu1996a}! Precision can only be gained in the rare cases where the vapor pressure of the probed molecule is large and accurately measured \cite{Sakurai1975a}. Our method however is applicable to molecular beams of low density and does not depend on the knowledge of vapor pressures.

On the other hand, our method compares favorably with the results of earlier depletion experiments~\cite{Wang1990a,Knickelbein1992a,Schmidt1999a} and electron detachment experiments \cite{Concina2008a}, exhibiting errors of the order of 10\%. 
This is due to the fact that it requires the absorption of on average less than a single soft
photon only. The photon does not need to cause any fragmentation,
ionization or other depletion. The probability of multi-photon absorption can be kept small, $P_{n\geq 2}(n_0) < 10\%$ for $n_0 < 0.5$, and
our spectroscopic method therefore probes most particles in the same state,
without any additional heating of the molecule or cluster.

\section{The effect of fluorescence}

One possible side effect still remains to be considered in our
discussion. Upon absorbing a photon, the molecules may either
redistribute the energy in an internal conversion process or reemit
the energy  as a fluorescence photon, typically on the time scale of
a few nanoseconds after the excitation. This can alter the reduction
of the interference contrast significantly, and it would be a major
source of error if it remained unconsidered. In fact, our exemplary
molecule H$_2$TPP does fluoresce with a quantum yield of about
$11\,\%$ \cite{Du1998a}.

A great advantage of our near-field setup is that we can separate
the side effects related to the recoil laser by varying the recoil
shift within one grating period. This is due to the strict
periodicity of the interference pattern, a feature absent in far
field interference. As an important application of that, we can thus
assess the effect of molecular fluorescence. In a simple model each
photon absorption is followed by an immediate fluorescence with a
probability given by the quantum yield $P_\text{fluo}$; otherwise
the excitation energy is deposited in the internal state of the
molecule. This assumes that the photon absorption rate during the
laser passage is small compared to the inverse of the typical
fluorescence time of the molecule, which is the case for typical
molecules and laser powers.

Each fluorescence emission can then be treated as a single
decoherence event that transforms the molecular Wigner function
$w(x,p)$ to $\int \diff q \, \widetilde{\gamma} (q) w(x,p-q)$,
corresponding to a probabilistic momentum kick with
distribution $\widetilde{\gamma} (q)$. The latter is given by the Fourier
transform of the decoherence function \cite{Hornberger2004a}
\begin{equation}
    \gamma (x) = \int_0^{\infty} \diff \omega \, F(\omega) \sinc \left( \frac{\omega x}{c} \right)
\, , \label{eqn:gamma}
\end{equation}
where $F(\omega)$ represents the normalized fluorescence spectrum.
By means of the probability $p_{n,k} = \binom{n}{k}
P_\text{fluo}^k (1-P_\text{fluo})^{n-k}$ of emitting $k$ out of
$n$ absorbed photons, the Fourier coefficients of the fringe
pattern can be expressed as
\begin{eqnarray}
    w_\ell (n_0,s) &=& w_\ell^{(0)} \exp \left[ - n_0 \left\{ 1 - \exp \left( \frac{2\pi i \ell s}{d}\right)  \right. \right. \nonumber \\
    &&\times \left. \left. \left( P_\text{fluo}\,\gamma \left( \frac{\ell \lambda_L s}{d} \right) + 1-P_\text{fluo} \right) \right\} \right].
\label{eqn:wlshift2}
\end{eqnarray}
In the absence of fluorescence, $P_\text{fluo}=0$, expression
(\ref{eqn:wlshift2}) reduces to Eq.~(\ref{eqn:wlshift}). The
absorption cross section $\sigma_\text{abs}$ (or equivalently the
mean number $n_0$ of absorbed photons, according to
Eq.~(\ref{eqn:n0})) and the quantum yield $P_\text{fluo}$ can then
be obtained one after the other by subsequently setting the recoil
shift to $s=d/2$ and to $s=d$, and by measuring the corresponding
reduction factors of the sinusoidal interference contrast,
\begin{eqnarray}
    \ln R^{-1} \left(n_0,\frac{d}{2}\right) &=& 2n_0 - P_\text{fluo} n_0 \left[1- \gamma \left( \frac{\lambda_L}{2} \right) \right], \label{eqn:lnR2} \\
    \ln R^{-1} (n_0,d) &=& P_\text{fluo} n_0 \left[1- \gamma \left( \lambda_L \right) \right]. \label{eqn:lnR3}
\end{eqnarray}
Since the latter depends only on the product $P_\text{fluo} n_0$ it
can be used to eliminate the fluorescence term in the first
reduction factor. The two reduction factors together thus determine
both the absorption cross section $\sigma_\text{abs}$ and the
quantum yield $P_\text{fluo}$. Note that this procedure requires
knowledge of the \emph{relative} fluorescence spectrum. This can for
instance be measured close to the molecular source where the density
is still high.

\begin{figure}
\includegraphics[width=8cm]{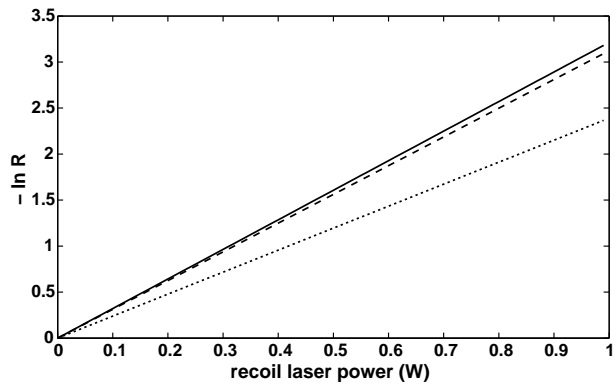}
\caption{\label{fig:fluo} Influence of fluorescence on the KDTLI
contrast reduction by the recoil laser. The theoretical
values of the negative logarithm of the contrast reduction factor
(\ref{eqn:lnR2}) are plotted versus the power of the recoil laser
for H$_2$TPP molecules at $v_z=175\,$m/s, with all experimental
parameters the same as in Fig.~\ref{fig:sigmafit}. The solid
line corresponds to the
case without fluorescence,  $P_\text{fluo}=0$, while the dashed line represents the
case of a realistic quantum yield, $P_\text{fluo}=11\,\%$, and the dotted line
corresponds to maximally possible value $P_\text{fluo}=100\,\%$.}
\end{figure}

Figure~\ref{fig:fluo} demonstrates the influence of fluorescence on
the recoil-induced contrast reduction for the H$_2$TPP example as
used in the simulation above.
We plot the negative natural logarithm of the reduction factor
as a function of the power of the recoil laser for three
instances: The solid line depicts the situation
(\ref{eqn:lnR1}) without fluorescence, as used in the simulation
in Fig.~\ref{fig:sigmafit}, while the dashed and the dotted line
represent the reduction factor (\ref{eqn:lnR2}) with a realistic
fluorescence quantum yield of $P_\text{fluo}=11\,\%$
\cite{Du1998a}, and with the maximum value $P_\text{fluo}=100\,\%$,
respectively. The underlying fluorescence spectrum is taken from
\cite{Du1998a}, which evaluates the decoherence function
(\ref{eqn:gamma}) as $\gamma(\lambda_L /2) = 0.49$. One observes in
Fig.~\ref{fig:fluo}
that the slope of the dashed line is $3\,\%$ smaller than the slope of the
solid line, which means that one would obtain a
value for the absorption cross section $\sigma_\text{abs}$ that is
$3\,\%$ too small if the
fluorescence effect was ignored. Larger quantum yields lead to a
more significant deviation of up to $25\,\%$, as illustrated by
the dotted line.

In principle, the visibility could be additionally reduced by the
recoil laser since the heating of the internal molecular state can
give rise to thermal emission and decoherence
\cite{Hornberger2004a,Hornberger2005a}. In fact, each absorbed laser
photon that is not reemitted by fluorescence increases the internal
temperature of the molecule by $\Delta T = h c /\lambda_L C$, with
$C$ the molecular heat capacity. 
However, on the time scale of the
proposed experiment the enhanced thermal radiation is expected to be
negligible for large molecules with capacities $C \gg k_B$, where the temperature increase per photon remains
typically below 100\,K for particles beyond 200\,atoms\cite{Hornberger2005a}. 
Such heating effects, but also a power dependence of
the absorption cross section, would manifest themselves in a
nonlinear behavior of the logarithmic contrast reduction when the
average number of absorbed photons is $n_0 > 1$.

Our method only measures the thermal average of the absorption cross
section at a temperature given by the source.
While in effusive sources
this temperature could be as high as several hundred Kelvin both
supersonic source and buffer gas cooled cluster sources would
prepare the particles in low internal states, often the vibrational
ground state. 

Nevertheless, the source temperature is of particular importance for small clusters
and molecules where internal heating can change the spectral
properties quite
dramatically~\cite{Ellert1995a,Chen1999a,Moseler2001a,Makarov2008a}.
While this appears to be a potential complication at first
sight the optical absorption cross section can actually be a valuable probe for the cluster
temperature as well~\cite{Schmidt1999a}. 
And future experiments
might explore this again systematically by starting from internally
cold particles which are subsequently heated using laser light.

\section{Conclusions}

Our analysis suggests that single photon recoil interactions within
a near-field matter wave interferometer are a promising tool for
absorption spectroscopy in molecular and cluster physics. The method
is robust and general since it does not depend on the
particularities of the interferometer, such as the details of the
molecular interaction with the diffracting element:  The
interferometric setup is only required to generate a fringe pattern
with a good modulation of the molecular density at a sufficiently
high signal-to-noise ratio.

The technique appears to be well-suited for a large variety of
neutral nanoparticles, including fluorescent molecules. It is
particularly applicable to particles with a strong internal state
conversion, a property found in many organic molecules and metal
clusters.

Quantum interference experiments planned for the near-future should
be capable to host a vast range of species in the mass range up to
10.000\,amu, while the principle can be easily scaled to larger
particles as soon as sufficiently slow and cold beams of massive
particles become available. A particular advantage of the present method is that one can extract
absorption cross sections from beams that are optically extremely
thin. Our simulation indicates that molecular beams with densities
down to $n\simeq 10^5\,{\rm cm}^{-3}$ should already suffice. For
the molecule H$_2$TTP used in the example above this corresponds to
an extinction length of about 65,000 km.

\section*{Acknowledgments}
We thank the Austrian science foundation FWF for support within
the projects SFB1505 and the doctoral
program W1210 (CoQuS), and the ESF EuroQUASAR program
MIME. K. H. is supported by the DFG within the Emmy Noether
program.

\end{document}